\documentclass[%
twocolumn,
superscriptaddress,
 amsmath,amssymb,
aps,
pra,
]{revtex4-2}

\newcommand {\tit}[1] {\textit{#1}}

\newcommand{\w}{\bar{\omega}}

\usepackage[normalem]{ulem}
\usepackage{amsmath}
\usepackage{amssymb}
\usepackage{esint}
\usepackage{graphicx}
\graphicspath{ {main_figs/} }
\usepackage{dcolumn}
\usepackage{bm}
\usepackage{braket}
\usepackage{breakurl}
\usepackage{seqsplit}
\usepackage[hidelinks,colorlinks]{hyperref}
\hypersetup{citecolor=[rgb]{0.25,0.14,0.63}}
\hypersetup{urlcolor=[rgb]{0.25,0.14,0.63}}
\hypersetup{linkcolor=black}

\newcommand{\pdn}[3]{\frac{\partial^#1 #2}{\partial #3^#1}}

\newcommand{\avg}[1]{\braket{#1}}

\begin{document}

\title{Thermalization of a quantum Newton's cradle in a one-dimensional quasicondensate}

\author{Kieran F. Thomas}
\affiliation{School of Mathematics and Physics, University of Queensland\text{,}  Brisbane\text{,} Queensland 4072\text{,} Australia }
\author{Matthew J. Davis}
\affiliation{ARC Centre of Excellence for Engineered Quantum Systems, School of Mathematics and Physics, University of Queensland\text{,}  Brisbane\text{,} Queensland 4072\text{,} Australia }
\author{Karen V. Kheruntsyan}%
\affiliation{School of Mathematics and Physics, University of Queensland\text{,}  Brisbane\text{,} Queensland 4072\text{,} Australia }

\date{\today}

\begin{abstract}
We study the nonequilibrium dynamics of the quantum Newton's cradle in a one-dimensional (1D) Bose gas in the weakly-interacting quasicondensate regime. This is the opposite regime to the original quantum Newton's cradle experiment of Kinoshita \textit{et al.} [Nature {\bf{440}}, 900 (2006)], which was realized in the strongly interacting 1D Bose gas. Using finite temperature \(c\)-field methods, we calculate the characteristic relaxation rates to the final equilibrium state. Hence, we identify the different dynamical regimes of the system in the parameter space that characterizes the strength of interatomic interactions, the initial temperature, and the magnitude of the Bragg momentum used to initiate the collisional oscillations of the cradle. In all parameter regimes, we find that the system relaxes to a final equilibrium state for which the momentum distribution is consistent with a thermal distribution. 
For sufficiently large initial Bragg momentum, the system can undergo hundreds of repeated collisional oscillations before reaching the final thermal equilibrium. The corresponding thermalization timescales can reach tens of seconds, which is an order of magnitude smaller than in the strongly interacting regime.
\end{abstract}

\maketitle

\section{Introduction}
The question of how isolated quantum systems relax after a disturbance \cite{rdyo07,rdo08} and reach a seemingly thermal equilibrium state \cite{no_relaxation} has been a topic of much interest in recent years (for reviews, see \cite{cr10,pssv11,yukalov11,efg15,ge16}). The experimental realization of a quantum Newton's cradle \cite{kww06}---a strongly interacting one-dimensional (1D) Bose gas undergoing repeated collisions in a harmonic trap--- was one of the first demonstrations of an isolated quantum system that did not thermalize over observable time scales, which were on the order of tens of seconds, corresponding to thousands of collisions. It was conjectured at the time that the lack of thermalization was due to the fact that the system is well approximated by an integrable Lieb-Liniger model \cite{ll63}. Such integrable systems are now understood to generally relax to non-thermal states described by a generalized Gibbs ensemble \cite{rdo08,r09,pssv11,kwe11,ck12,legrskrmgs15}, which puts further constraints on the system dynamics compared to those present in generic (nonintegrable) systems described by conventional ensembles of statistical mechanics.

On the other hand, an experiment performed by Hofferberth \tit{et al.} \cite{hlfss07}, studying the relative phase dynamics of a coherently split 1D quasicondensate \cite{petrov2000}, \textit{i.e.}, a phase fluctuating 1D Bose gas in the weakly interacting regime, have initially suggested that these dynamics relax over millisecond time scales. This is orders of magnitude smaller than in the quantum Newton's cradle, despite the fact that the strictly 1D quasicondensates can be well approximated by the same integrable Lieb-Liniger model, with the proviso that the harmonic trapping potential breaks the exact integrability (just as it does in the strongly interacting Newton's cradle case). Hence, it was a question of interest whether the stark difference in characteristic relaxation times in these two experiments was caused by the proximity of the system to the integrable regime, or the specifics of the dynamical scenario being considered. 

Since these early experiments, there has been significant progress in the theoretical understanding of these seemingly contradictory results. The fast relaxation to a seemingly thermal state in the experiments with split quasicondensates can be understood as a result of purely phase dynamics and the phenomenon of prethermalization: the system dephases to a prethermalized state on a short time scale, where certain observables (in this case interference contrast \cite{gritsev2006full}) of a nonequilibrium, long-lived transient state become indistinguishable from those of a thermal equilibrium state \cite{gklkrsmsds12,kitagawa2011dynamics,smith2013prethermalization}. Relaxation to a final equilibrium  state {at a higher final temperature than the pre-thermalised state} 
takes much longer and depends on the evolution of the quasicondensate density \cite{gklkrsmsds12,smith2013prethermalization}.  The splitting of the initial equilibrium quasicondensate here creates two initially phase-coherent samples, which are in a nonequilibrium (excited) state of the underlying trapping potential, and while the relative phase relaxes rather quickly, the density profiles of the two quasicondensates start to undergo breathing-mode oscillations which take much longer to damp out. A single 1D quasicondensate undergoing damping of this kind, excited by an alternative but equivalent \cite{KKMD} confinement quench, has been observed experimentally in Ref.~\cite{Fang14} and was also evident in the theoretical study of Ref.~\cite{bsdk16}. However, a detailed understanding of such damping is yet to be developed. {We further note that even without excitation of the breathing-mode oscillations, which can be suppressed by an
adequate optimum control protocol, the work of Stimming \textit{et al.} \cite{Stimming} indicates that the temperature of the final true equilibrium state will likely differ from that of the intermediate pre-thermalised state due to the sub-exponential decay of coherence between the split quasicondensates.}

The question of how long would it take a strictly 1D quasicondensate, \textit{i.e.}, a weakly- rather than strongly-interacting 1D Bose gas, to relax in the quantum Newton's cradle setting of Ref.~\cite{kww06} remains unanswered. Here we address this question by performing a numerical experiment of the quantum Newton's cradle setup in a finite-temperature 1D quasicondensate in the quantum degenerate regime. The relaxation in this scenario involves both  phase and density dynamics right from the onset of oscillations, unlike the split quasicondensate case. We find that such a quasicondensate can take several seconds (corresponding to hundreds of oscillations) to thermalize, which is somewhat shorter compared to the equivalent experiment in the strongly interacting regime, but still much longer  than relaxation due to purely phase dynamics in the split quasicondensate. {We emphasise that this estimate is based on purely one-dimensional dynamics governed by the 1D Gross-Pitaevskii equation (GPE) in the classical $c$-field approach.  It ignores the effect of interactions on the radial width of the quasicondensate, which could potentially be taken into account via an effective 1D nonpolynomial GPE of Ref.~\cite{Salasnich}, although the validity of applying a c-field approach to this equation of motion is unclear. Any significant radial dynamics would likely further reduce the thermalization time scales compared to those reported here.}

{While the long relaxation times of the original Newton's cradle experiment can indeed be understood by the proximity of the system to integrability, the said proximity here is two-fold. The first is the proximity to a uniform 1D Bose gas that can be described by the integrable Lieb-Liniger model~\cite{ll63} of bosons interacting via a pairwise contact interaction. Even though the harmonic trap (in which the two momentum components of the gas repeatedly collide) breaks this integrability \cite{M11}, the breakdown can be considered weak because the longitudinal trapping is relatively weak. Accordingly, the apparent lack of relaxation to a conventional thermal state (\textit{i.e.}, one described by a conventional grand-canonical ensemble) can be understood as a quantum analog of the Kolmogorov-Arnold-Moser (KAM) theorem~\cite{pssv11,Caux_KAM_2015} and the approximate applicability of the generalized Gibbs ensemble. The second aspect of the proximity to the integrability point is that the experiments of Ref.~\cite{kww06} were performed in the strongly interacting regime, which is closer to the Tonks-Girardeau gas of impenetrable (or hard-core) bosons \cite{Girardeau1960}. The strictly Tonks-Girardeau gas corresponds to the limit of infinitely strong interactions of the Lieb-Liniger model and is integrable even in a harmonic trap \cite{Girardeau_Bose_Fermi,Girardeau2000,yukalov2005fermi,agbk16} (theoretical modelling of the quantum Newton's cradle in the Tonks-Girardeau regime can be found in Refs.~\cite{Caux-QNC,agbk16}). Therefore, a harmonically trapped 1D Bose gas with large but finite interaction strength (which is the case in practice) is closer to being integrable in this sense compared to a similarly trapped but weakly-interacting 1D quasicondensate. Accordingly, such a gas might be expected to relax on longer time scales than a quasicondensate, which is what our results demonstrate.}

Additional practical matters can compromise integrability and affect relaxation pathways and time scales. Such factors include deviations of interparticle interactions from point-like character, deviations of the trapping potential from being purely harmonic or truly 1D \cite{kww06,Schmiedmayer_cradle_2018}, and virtual transverse excitations that can lead to effective three-body collisional relaxation. For example, Refs.~\cite{mss08,ms10} show that such virtual three-body collisions result in dramatically different relaxation rates in the weakly- and strongly-interacting quasi-1D Bose gases, hence providing an alternative explanation of the comparatively shorter relaxation times in the experiments with split quasicondensates. In light of this situation, the question of how 1D quasicondensates in a quantum Newton's cradle setup relax in a harmonic trap is a different scenario to those considered previously.
It is also within reach of being addressed experimentally \cite{bvw09}, hence the motivation for the present work.

Finally, we would like to mention some recent closely related work by Bland~\emph{et al.}~\cite{Bland_2018}, who have also studied the integrability of the weakly-interacting 1D Bose gas in a harmonic trap with the Gross-Pitaevskii equation. Even though Bland~\emph{et al.} conclude that the harmonically trapped 1D Bose gas in the GPE regime is quasi-integrable and hence does not relax to a thermal state, we show here that this conclusion does not hold true for effective nonlinearities larger than the ones considered in their work. We also note the recent work by Bastianello~\emph{et al.}~\cite{Bastianello_2020} who consider the relaxation of a 1D harmonically trapped Bose gas using a generalized hydrodynamics approach, {which has a different conceptual foundation to our approach.  They also find that the system thermalizes at long times, giving us further confidence in our conclusions.}

This paper is organized as follows. In Sec.~\ref{techniques}, we describe the $c$-field method used in the numerical simulations of the quantum Newton's cradle in a 1D quasicondensate, and parametrize the system in terms of the relevant dimensionless parameters. In Sec.~\ref{examples}, we present examples of relaxation dynamics in three typical regimes. {In Sec. \ref{sec:power}, we analyze the power spectrum of the population of the lowest energy harmonic oscillator mode for the quantum Newton's cradle simulations, as in Ref.~\cite{Bland_2018}.} In Sec.~\ref{phase-diagram}, we discuss simple physical considerations required for observing persistent collisional oscillations in certain parameter regimes. In Sec.~\ref{decay}, we characterize the system dynamics in terms of relaxation rates and identify the three different dynamical regimes of the system in the relevant space of dimensionless parameters. Finally, in Sec.~\ref{conclusion}, we summarize our findings and discuss their implications.

\section{The $C$-field method for simulating the quantum Newton's cradle}
\label{techniques}

The $c$-field (or classical field) method \cite{Castin:2000,Davis:2001b,bbdbg08} is a widely used approach to studying equilibrium and nonequilubrium properties of finite temperature Bose gases, including a weakly-interacting 1D Bose gas in the quasicondensate regime \cite{bsdk16}. 
The essence of the method is to treat the low-energy coherent band of the quantum Bose field as a classical field $\psi_C(x,t)$, thus ignoring all quantum fluctuations and the discrete nature of the particles that make up the field. In this approach, finite-temperature equilibrium configurations are sampled by integrating the stochastic projected Gross-Pitaevskii equation (SPGPE) for sufficiently long times that the memory of the initial trial state is lost. These are then used to establish the initial state of a harmonically trapped 1D quasicondensate. The subsequent real-time dynamical evolution, which is set by applying a Bragg pulse that initiates the Newton's cradle collisions, is then described by evolving an ensemble of individual realizations from the SPGPE according to the projected Gross-Pitaevskii equation (GPE).

\subsection{Parameterization of the model}

To write down the $c$-field equations using an efficient parametrization, we recall from the works of Lieb-Liniger \cite{ll63} and Yang-Yang \cite{yy69} that the properties of a uniform 1D Bose gas, with linear (1D) density $\rho$ and temperature $T$, can be completely characterized by two dimensionless parameters \cite{kgds03,kgds05}: the dimensionless interaction strength $\gamma=m g/(\hbar^2 \rho)$ and dimensionless temperature ${\mathcal{T}}=2\hbar^2 k_B T/(mg^2)$, where $m$ is the mass of the particles, $k_B$ is the Boltzmann constant, and $g$ is the 1D interaction strength, which can be expressed via the 3D scattering length $a$ and the frequency of the transverse harmonic potential $\omega_{\perp}$ as $g\approx2\hbar\omega_{\perp}a$ far away from confinement induced resonances \cite{o98}. 

In this paper, we restrict ourselves to the study of a weakly-interacting ($\gamma\ll 1$) 1D Bose gas in a phase-fluctuating quasicondensate regime. This corresponds to a quantum degenerate gas in which the interparticle interactions play a nonperturbative role. The observable that we use to characterize the relaxation dynamics is the momentum distribution of the gas. Accordingly, as long as the (low-momentum) bulk of the momentum distribution is concerned, the quasicondensate regime is defined by the condition $g\rho e^{-2\pi/\sqrt{\gamma}}\ll k_BT\ll \sqrt{\gamma}\hbar^2\rho^2/m$ \cite{Bouchoule-mom-corr,condition-position-space}. At these temperatures, the highly-occupied, low momentum modes are dominated by thermal fluctuations rather than vacuum fluctuations, ensuring the applicability of the $c$-field method. Vacuum fluctuations become important at temperatures $k_BT\ll g\rho e^{-2\pi/\sqrt{\gamma}}$, which are exponentially small in the regime of weak interactions, $\gamma \ll 1$, and are beyond the reach of current ultracold atom experiments. Thus, for all practical purposes, the quasicondensate regime corresponds to $k_BT\ll \sqrt{\gamma}\hbar^2\rho^2/m$ \cite{IBG}. Moreover, as discussed in Refs.~\cite{Castin:2000,Bouchoule-mom-corr,bsdk16}, in the $c$-field approach the system (a uniform quasicondensate) can be completely characterized by a single dimensionless parameter, $\chi=\frac{1}{2}\gamma^{3/2} {\cal{T}}$, which is a combination of $\gamma$ and  ${\cal{T}}$ and which satisfies $\chi\ll 1$ (according to $k_BT\ll \sqrt{\gamma}\hbar^2\rho^2/m$).

For a harmonically trapped (nonuniform) 1D quasicondensate with longitudinal potential $V(x)=\frac{1}{2}m\omega^2x^2$, one needs an additional parameter---the trap frequency $\omega$---to completely characterize the system. In this nonuniform case, the linear density becomes position dependent and describes the density profile of the gas $\rho(x)$. Accordingly, the dimensionless interaction strength also becomes position dependent, $\gamma(x)= m g/(\hbar^2 \rho(x))$, while the dimensionless temperature ${\cal{T}}$ continues to serve as the global equilibrium temperature of the system. For a given chemical potential $\mu$, which fixes the total number of particles $N$ in the system, the density profile $\rho(x)$ is unique. Therefore the peak density $\rho_0\equiv \rho(0)$ in the trap centre $x=0$ can be used to define a dimensionless interaction strength $\gamma_0=m g/(\hbar^2 \rho_0)$ that plays the role of a global interaction parameter for the trapped system.

The essence of the efficient parametrization of the trapped 1D quasicondensate lies in the fact that, apart from the dependence on the longitudinal trapping frequency $\omega$, it can still be completely characterized by a single dimensionless parameter, which is a combination of the dimensionless interaction strength in the trap center, $\gamma_0$, and the global dimensionless temperature $\mathcal{T}$:
\begin{align}
\chi_0=\frac{1}{2}\gamma_0^{3/2} {\cal{T}}.
\label{eq:chi_0}
\end{align}
Explicitly, $\chi_0$ is given by $\chi_0\equiv k_BT/[\hbar\rho_0\sqrt{g\rho_0/m}]$; similar to \(\chi\) it must satisfy $\chi_0\ll 1$ in the quasicondensate regime; and we note that its definition is related to the dimensionless parameters $A$ and $\chi^{[C]}$, used, respectively, in Refs.~\cite{bsdk16,Castin:2000} (see also \cite{Bouchoule-mom-corr}). In particular, in Ref.~\cite{Castin:2000}, $\chi^{[C]}$ [see, Eq.~(77)] is defined in terms of a product of ratios of characteristic energy- and length-scales of the uniform problem, $\chi^{[C]}\!=\!\frac{1}{2\pi}(\frac{g\rho}{k_BT})(\rho\lambda)^2$, where $\lambda\!=\!\sqrt{2\pi\hbar^2/mk_BT}$ is the thermal de Broglie wavelength, and our $\chi$ is related to $\chi^{[C]}$ via $\chi\!=\!1/\sqrt{\chi^{[C]}}$, whereas our $\chi_0$ is defined similarly but in terms of the peak density $\rho_0$.

\subsection{Dimensionless $c$-field equations in 1D}

To arrive at the dimensionless form of the SPGPE and GPE using this parametrization, we introduce the dimensionless coordinate $\xi=x/x_0$, time $\tau=t/t_0$, and dimensionless field $\varphi_C(\xi,\tau)=\psi_C(x,t)/\psi_0$, using the respective length-, time-, and field-scales introduced according to:
\begin{align}
x_0 & =  \frac{\hbar^{4/3}}{m^{2/3}g^{1/3}(k_BT)^{1/3}}, \\ \label{eq:lengthscale}
t_0 & =  \frac{m x_0^2}{\hbar} = \frac{\hbar^{5/3}}{m^{1/3}g^{2/3}(k_BT)^{2/3}}, \\
\psi_0 & = \frac{\hbar}{x_0 \sqrt{mg }} =  \left( \frac{m(k_BT)^{2}}{\hbar^2g}\right)^{1/6}.
\end{align}

With these scaled variables, the dimensionless time-dependent GPE reads as
\begin{equation}
\frac{\partial \varphi_C(\xi,\tau)}{\partial{\tau}}=-i\mathcal{L} \, \varphi_C(\xi,\tau),
\label{GPE}
\end{equation}
where the nonlinear operator $\mathcal{L}$ is defined via
\begin{equation}
\mathcal{L} \equiv  -\frac{1}{2} \pdn{2}{}{\xi} + |\varphi_C|^2 + \frac{1}{2} \bar{\omega}^2 \xi^2,
\end{equation}
whereas the dimensionless SPGPE reads as
\begin{equation}
d \varphi_C(\xi,\tau) \!= \! [-i \mathcal{L} + \bar{\kappa}_{th} (\bar{\mu}\!-\!\mathcal{L})] \varphi_C d \tau + \sqrt{2 \bar{\kappa}_{th} } d \overline{W}.
\label{eqn:spgpe_thermal}
\end{equation}
Here, $\bar{\omega}=\omega t_0$ is the dimensionless trap frequency, $\bar{\mu}=\mu t_0/\hbar$ is the dimensionless chemical potential and \(\bar{\kappa}_{th} = \hbar \kappa_{th}\) is the rescaled growth rate  with a numerical value that has no consequence for the final equilibrium configurations and can be chosen for numerical convenience \cite{bsdk16}. Additionally, \(d\overline{W}\) is a complex delta-correlated noise satisfying \(\avg{d\overline{W}^*(\xi,\tau)d\overline{W}(\xi',\tau)}=\delta(\xi-\xi')d\tau\). We point out that 
even though we are referring to Eq. (\ref{eqn:spgpe_thermal}) as the SPGPE, 
for the results presented here the actual projection operator, which sets up the high-energy cutoff for the classical field region, is effectively imposed by the choice of numerical grid for simulations. Apart from this, we have performed a range of simulations in the harmonic oscillator basis with a strictly controlled cutoff, and found that our results are essentially unchanged. Due to the strictly 1D nature of the problem at hand, the actual cutoff dependence of the dynamics is very weak (see Ref.~\cite{bsdk16} for further details).

In the above dimensionless form, the nonlinearity constant in the GPE is always equal to unity, and the normalization condition that gives the total number of particles in the system reads $N=\int |\psi(x,t)|^2dx=\psi_0^2x_0 \bar{N}$, where $\bar{N}=\int |\varphi(\xi,\tau)|^2 d \xi$.

In the Thomas-Fermi (TF) limit of an inverted parabolic density profile, the chemical potential of a harmonically trapped quasicondensate is given by $\mu=g\rho_0$, and thus the dimensionless chemical potential $\bar{\mu}$ can be expressed in terms of, and interchanged with, the earlier introduced dimensionless parameter $\chi_0$ as
\begin{equation}
\bar{\mu}=\chi_0^{-2/3}. \label{eqn:mu_bar}
\end{equation} 
Given that $\rho_0\!=\!(9m\omega^2N^2/32g)^{1/3}$ in the TF limit, $\chi_0$ itself can be expressed as $\chi_0\!=\!4\sqrt{2}k_BT/(3N\hbar\omega)$ [using $N\!=\!\psi_0^2x_0 \bar{N}$ and $\bar{N}\!=\!4 \sqrt{2}\bar{\mu}^{3/2}/(3\bar{\omega})\!=\!4 \sqrt{2} /(3\chi_0\bar{\omega})$]. Beyond the TF limit, the dimensionless chemical potential $\bar{\mu}$ can still be traded off with $\chi_0$ as an input parameter, with the understanding that the simple relationship $\mu=g\rho_0$ between $\mu$ and the peak density $\rho_0$ is now only approximate, whereas the exact relationship has to be determined numerically \emph{a posteriori}.

Thus, a harmonically trapped 1D quasicondensate can be completely characterized by just two dimensionless parameters, $\bar{\mu}$ and $\bar{\omega}$, or by $\chi_0\ll 1$ and $\bar{\omega}$. Each choice of $\chi_0$ can be realized with a range of values of the dimensionless interaction $\gamma_0$ and temperature $\mathcal{T}$ according to Eq.~(\ref{eq:chi_0}). The condition that the density profile is well approximated by the TF inverted parabola, $R_{TF}\!\gg \!l_{ho}$, is equivalent to $\bar{\omega}\!\ll \!\bar{\mu}$. This implies that for each choice of $\chi_0\ll 1$, the dimensionless trap frequency $\bar{\omega}$ must further satisfy $\bar{\omega}\!\ll \!\chi_0^{-2/3}$.
Furthermore, $\bar{\omega}$ can be expressed via the ratio of the length scale $x_0$ and the harmonic oscillator length $l_{ho}=\sqrt{\hbar/m\omega}$ as
\begin{equation}
\bar{\omega}=\left(\frac{x_0}{l_{ho}}\right)^2, 
\end{equation}
or explicitly $\bar{\omega}= \hbar^{5/3}m^{1/2}\omega/[g^{2/3}(k_BT)^{2/3}]$. Therefore, rewriting this as $\bar{\omega}^{-3/2}=mgk_BT/(m^{3/2}\omega^{3/2}\hbar^{5/2})$, one can see that for a specific atomic species (with given $m$ and $a$) and any particular choice of the trap frequencies $\omega$ and $\omega_{\perp}$ (the latter setting the value of the coupling constant $g=2\hbar\omega_{\perp}a$), fixing the temperature of the gas $T$ is equivalent to fixing the value of $\bar{\omega}$ and vice versa. With $\bar{\omega}$ constant, the value of the dimensionless parameter $\chi_0$ is now governed only by the choice of the total number of atoms $N$ or equivalently the peak density $\rho_0$.

\subsection{Choice of simulation parameters}

For the numerical simulations we have chosen $\bar{\omega}=0.0696$ and varied $\chi_0\in [0.01, 0.1]$, a range corresponding from deep within the thermal quasicondensate to a near degenerate ideal Bose gas regime (see Refs.~\cite{kgds05,Sykes:2008,jabkb11} for further details on the regimes of an interacting 1D Bose gas). The lower (upper) bound in $\chi_0$ corresponds to higher (lower) atom number $N$ and hence to conditions that are effectively deeper in the low-temperature quasicondensate regime (or are further away from the degenerate ideal Bose gas regime). Considering $^{87}$Rb atoms ($m\simeq1.44\times10^{-25}$ kg, $a\simeq 5.31$ nm) in a harmonic trap of frequency $\omega/2\pi=3$ Hz as an example, these choices of dimensionless parameters can be realized with $\omega_{\perp}/2\pi=6$ kHz, $T=13$ nK, and by varying the atom number $N\in [8.5\times 10^{2}, 8.5 \times 10^{3}]$, which is typical of current experiments on 1D quasicondensates. Note that while this is just one choice of dimensionless parameters, the same $\bar{\omega}$ and $\chi_0$ can be realized with other combinations of trap frequencies, temperatures, atom numbers, and atomic species.

\begin{figure*}[ht!]
\includegraphics[width=17.2cm]{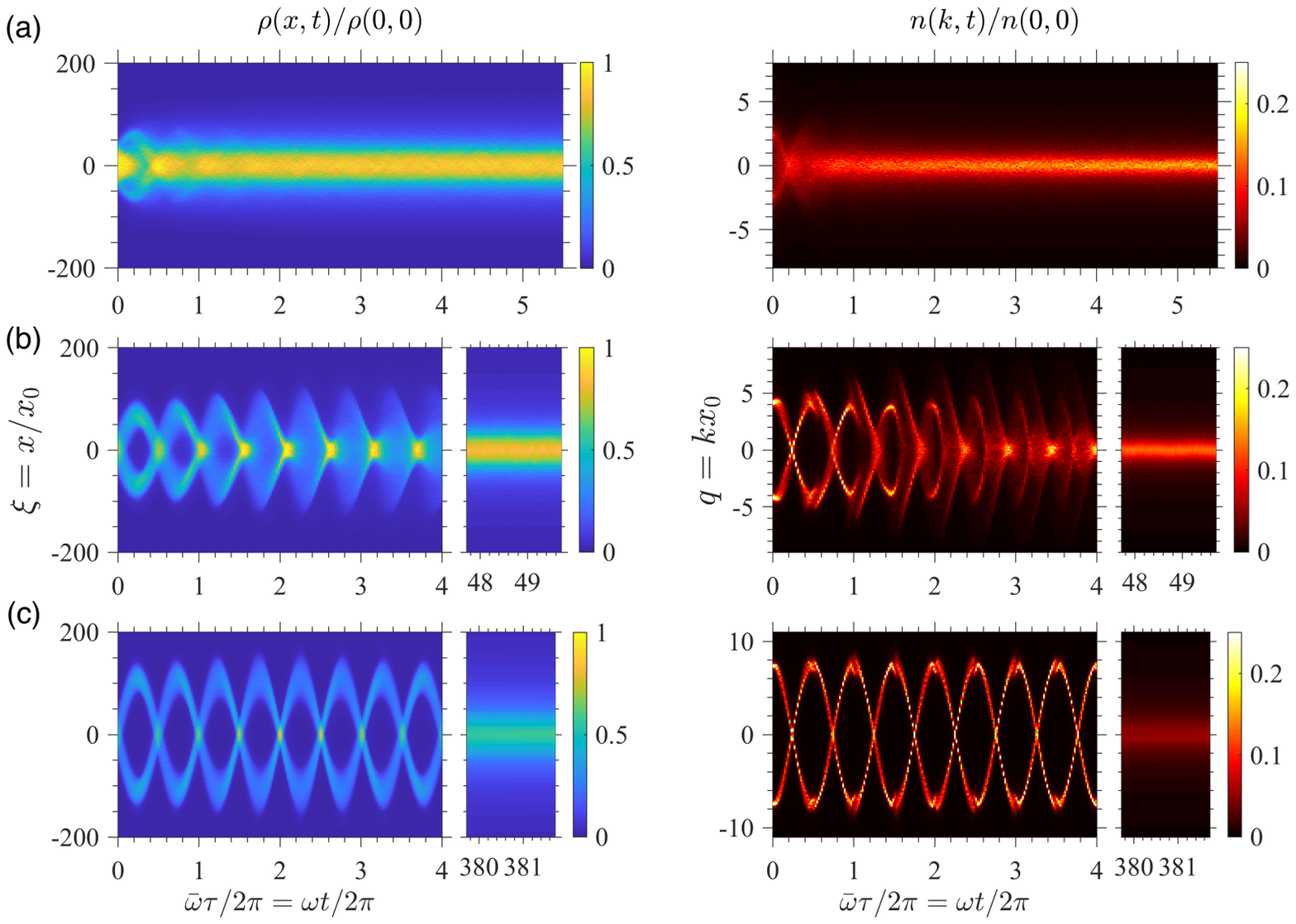}
\caption{Evolution of the position-space (left) and momentum-space (right) density distributions in the quantum Newton's cradle setup of a harmonically trapped 1D quasicondensate, generated from $c$-field simulations. The three examples displayed all have parameter values of \(\bar{\omega}=0.0696\) and \(\chi_0 = 0.0562\) for the initial thermal equilibrium state, whereas the wavenumber for the Bragg pulse varies as follows: (a) $q_0=1.16$; (b) $q_0=2.06$; and (c) $q_0=3.67$.
}
\label{fig:mom-distr}
\end{figure*}

To initiate the Newton's cradle dynamics, we model the application of a sequence of Bragg pulses \cite{wwdp05} which split the initial equilibrium quasicondensate into two counter-propagating wavepackets with momenta $\pm 2\hbar k_0$ corresponding to the lowest diffraction orders of Bragg scattering. Ideally, this initializes the $c$-field configuration denoted $\psi_C(x,t\!=\!0^+)$ into a superposition of the form 
\begin{align}
\psi_C(x,t\!=\!0^+) &= \frac{\psi_C(x,t=0^-)}{\sqrt{2}} \left( e^{i 2k_0 x}+e^{-i 2k_0 x} \right),
\label{Bragg}
\end{align}
where \(\psi_C(x,t=0^-)\) is the finite temperature equilibrium configuration. This state is then evolved in real time according to the GPE. This idealized form of the Bragg pulse is a good approximation to a realistic experimental implementation of a sequence of Bragg pulses tuned to operate in the Bragg regime \cite{wwdp05}, which corresponds to the condition that the Brag momentum  \(2\hbar k_0\) is much greater than the width of the momentum distribution of the quasicondensate. A comparison of the numerical implementations of the above idealized Bragg pulse and the more realistic pulse presented in Ref.~\cite{wwdp05} showed small differences in our results during the initial collisional oscillation cycles (see also the discussion of Figs.~\ref{fig:nm} and \ref{fig:zero_peak} below), however, no appreciable differences were found in the approach to the final relaxed state for all parameter regions.

In our simulations, we considered momentum kicks of $k_0\in [10^{6}, 10^{7}]$ m$^{-1}$. For $^{87}$Rb atoms and other relevant parameter choices as above (namely, $\omega_{\perp}/2\pi=6$ kHz and $T=13$ nK), this corresponds to a dimensionless momentum $q_0=k_0x_0$ in the range $q_0\in [1.16, 11.6]$. In harmonic oscillator units, $q_0$ can be converted to a dimensionless momentum $q_0^{(ho)}\equiv k_0l_{ho}=q_0/\sqrt{\bar{\omega}}$, and for $\bar{\omega}=0.0696$ this range of $q_0$ would correspond to $q_0^{(ho)}\in [4.4, 44]$.
From the practical point of view, considering momentum kicks significantly beyond the considered range of $q_0\in [1.16, 11.6]$ would be either of no physical interest in terms of producing (for smaller $q_0$) the collisional dynamics that we are interested in, or would be computationally too demanding (for larger $q_0$) in terms of the numerical grids required to capture the relevant physics at high momenta with sufficient resolution in both the momentum and position spaces.

\section{Results and discussion}
\label{results}

\subsection{Examples of relaxation dynamics}
\label{examples}

Typical examples of $c$-field simulations of the quantum Newton's cradle for a 1D quasicondensate are illustrated in Fig.~\ref{fig:mom-distr}, where we show the relaxation dynamics of the real-space density profile and the momentum distribution of the gas following the Bragg pulse. 
Figure~\ref{fig:mom-distr}(a) represents an example of fast relaxation, occurring in the regime where the clouds are extremely wide compared to the maximum spatial separation due to the weak momentum kick, and hence are always overlapped to some extent. This causes the system to relax without developing any appreciable collisional oscillations. Figure~\ref{fig:mom-distr}(b) is an example in the intermediate quasi-periodic regime with a stronger momentum kick; the clouds separate completely, but the momentum kick is still not large enough to lead to persisting collisions and periodic oscillations. Finally, Fig.~\ref{fig:mom-distr}(c) illustrates an example in the periodic regime, with the momentum kick larger than in (b); the collisional oscillations persist  in this system with no noticeable damping for many (tens to a hundred) periods, and take over a hundred oscillations to fully dampen.

\begin{figure}[tbp]
\includegraphics[width=8.5cm]{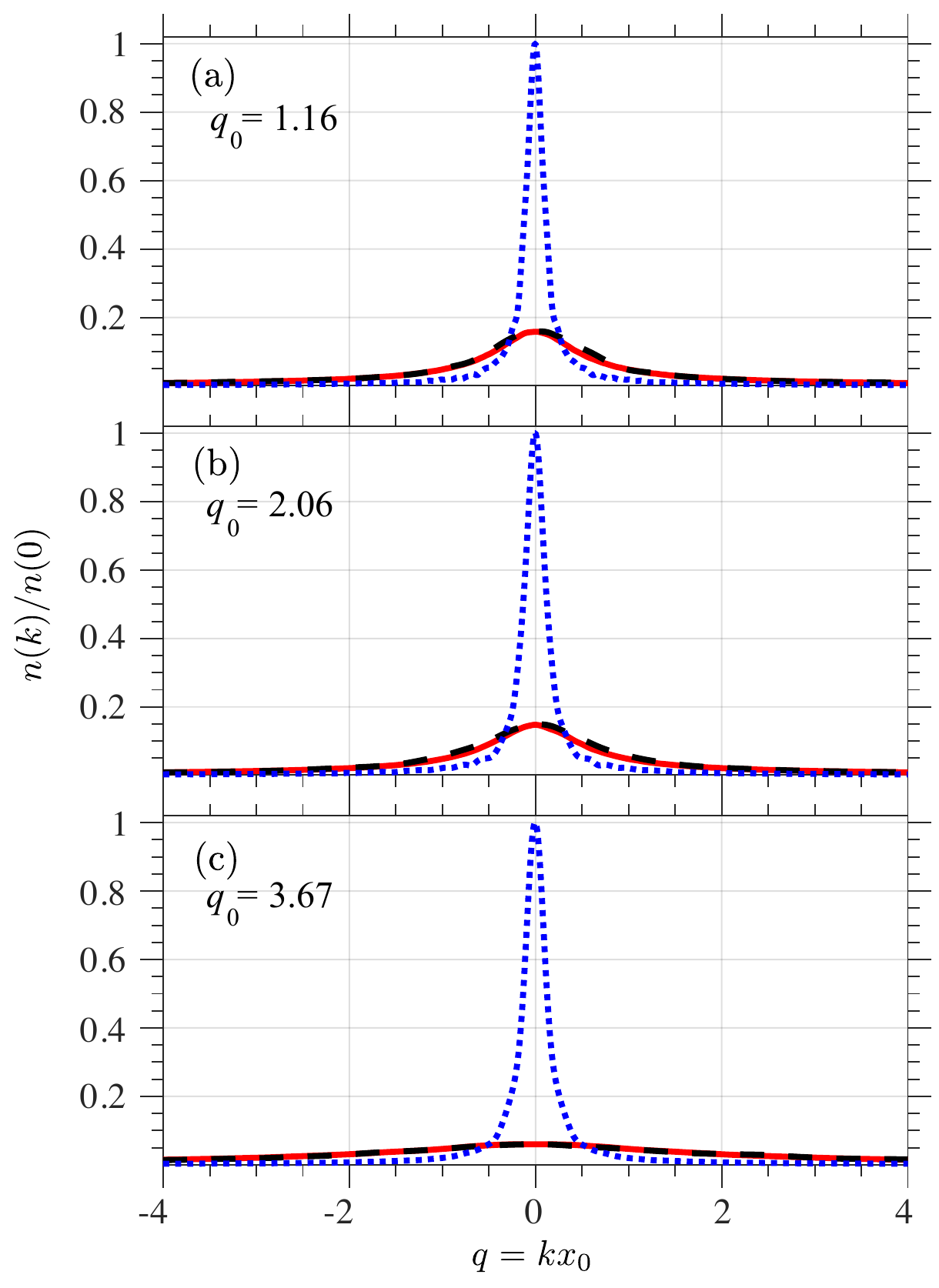}
\caption{Momentum distributions of a harmonically trapped 1D quasicondensate in the quantum Newton's cradle setup. In each example, we show the initial (blue, dotted line), relaxed (red, full), and thermal fitted distributions (black, dashed). The initial and relaxed distributions displayed in (a),
(b), and (c) are, respectively, for the same parameters as in Fig.~\ref{fig:mom-distr}, \emph{i.e.}, for $(\chi_0,\bar{\omega)=(0.0562, 0.0696)}$ initially and three different values of $q_0$ as shown, whereas the thermal fitted distributions, plotted for comparison with the relaxed ones, are for the following fitting parameters: (a) $(\chi_0,\bar{\omega})=(0.250, 0.045)$; (b) $(\chi_0,\bar{\omega})=(0.238,0.0462)$, and (c) $(\chi_0,\bar{\omega})=(0.500,0.150)$.
}
\label{fig:fits}
\end{figure}

\begin{figure}[tbp]
\includegraphics[width=8.55cm]{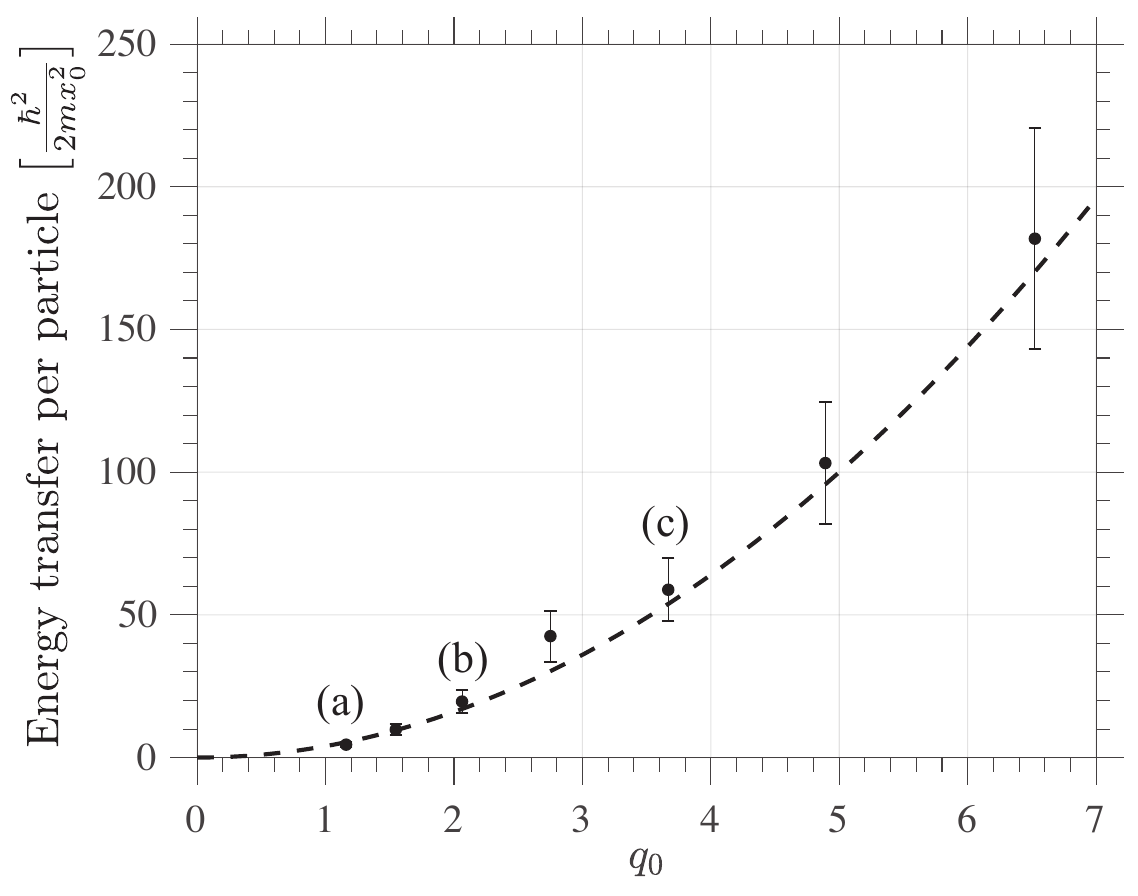}
\caption{Comparison of the total internal energy difference per atom (circles) between the initial and relaxed states with the kinetic energy per atom (dashed line) of $\hbar^2(2k_0)^2/2m$ (or $4q_0^2$, in units of  $\hbar^2/(2mx_0^2)$), transferred by the symmetric Bragg momentum kick of $\pm 2\hbar k_0$, as functions of the dimensionless momentum $q_0$, for \((\chi_0,\w)=(0.0562,0.0696)\). The data points represent the numerically calculated values of the average internal energy, averaged over 128 stochastic realizations, with the error bars indicating one standard deviation from the average; the points labelled (a), (b), and (c) correspond to the examples of Fig.~\ref{fig:mom-distr}. All energies have been scaled by $\hbar^2/(2mx_0^2)$; when multiplied by the total number of atoms $N$, this translates to $N\hbar^2/(2mx_0^2)=\frac{1}{2}k_BT\bar{N}$, where $\bar{N}=\int |\varphi(\xi,\tau)|^2 d \xi =4 \sqrt{2} /(3\chi_0\bar{\omega})$ and $T$ is the initial temperature. 
}
\label{fig:trans}
\end{figure}

In Fig.~\ref{fig:fits} we show the momentum distributions of the initial (\(t=0^-\)) and the final relaxed states for the examples of Fig.~\ref{fig:mom-distr}, and compare the final distributions with those of an equilibrium thermal state at a certain (higher) temperature. The best-fit thermal state, in which the temperature serves as a fitting parameter, is generated using the same SPGPE equation as the one used to initialize the quantum Newton's cradle, Eq.~(\ref{eqn:spgpe_thermal}). As we see, in all these examples the quantum Newton's cradle eventually relaxes to a thermal like state at a higher temperature (furthermore, this observation holds for all other tested parameter combinations). This conclusion is further supported by the observation (see Fig.~\ref{fig:trans}) that the increase in the internal energy of the relaxed state relative to the initial state generally matches the total kinetic energy added to the system by the Bragg pulse.

\subsection{Power spectrum}
\label{sec:power}

{In this subsection we perform a power spectrum analysis of the dynamics of the population of the lowest energy harmonic oscillator mode for the quantum Newton's cradle examples of Fig.~\ref{fig:mom-distr} in order to connect our results with related work by Bland~\emph{et al.}~\cite{Bland_2018}.

Bland~\emph{et al.}~\cite{Bland_2018}~have also studied the integrability of the weakly-interacting 1D Bose gas in a harmonic trap within the GPE approximation, in addition to a suitably truncated expansion over harmonic oscillator eigenmodes also known as the Galerkin approximation. More specifically, Bland~\emph{et al.}\ considered a small mode number Galerkin expansion of the GPE mean-field in both a hard-wall box potential and a harmonic potential, and compared the results of the time-dependent integration within this approximation to those of the full 1D GPE integration for these two situations. They considered a scenario where the lowest $M=16$ modes are populated with random phases and amplitudes in the two traps, which mimics the effect of thermal fluctuations in a finite-temperature system (here, essentially an infinite temperature sample).  They then analyzed the power spectrum of the Galerkin expansion coefficient of the lowest energy mode. They found that the harmonic trap power spectra was quasi-discrete, ``with no visible continuous component'', whereas the box trap spectra gave a continuous spectrum. These findings suggest a ``slow onset of chaotization''~\cite{Bland_2018} and a lack of ``a necessary
manifestation of ergodicity''~\cite{Bland_2018} in the harmonic trap. On this basis, Bland~\emph{et al.} concluded that a harmonically trapped, weakly-interacting 1D Bose gas in the GPE regime is quasi-integrable and is not ergodic.

\begin{figure}[tbp]
\includegraphics[width=8.5cm]{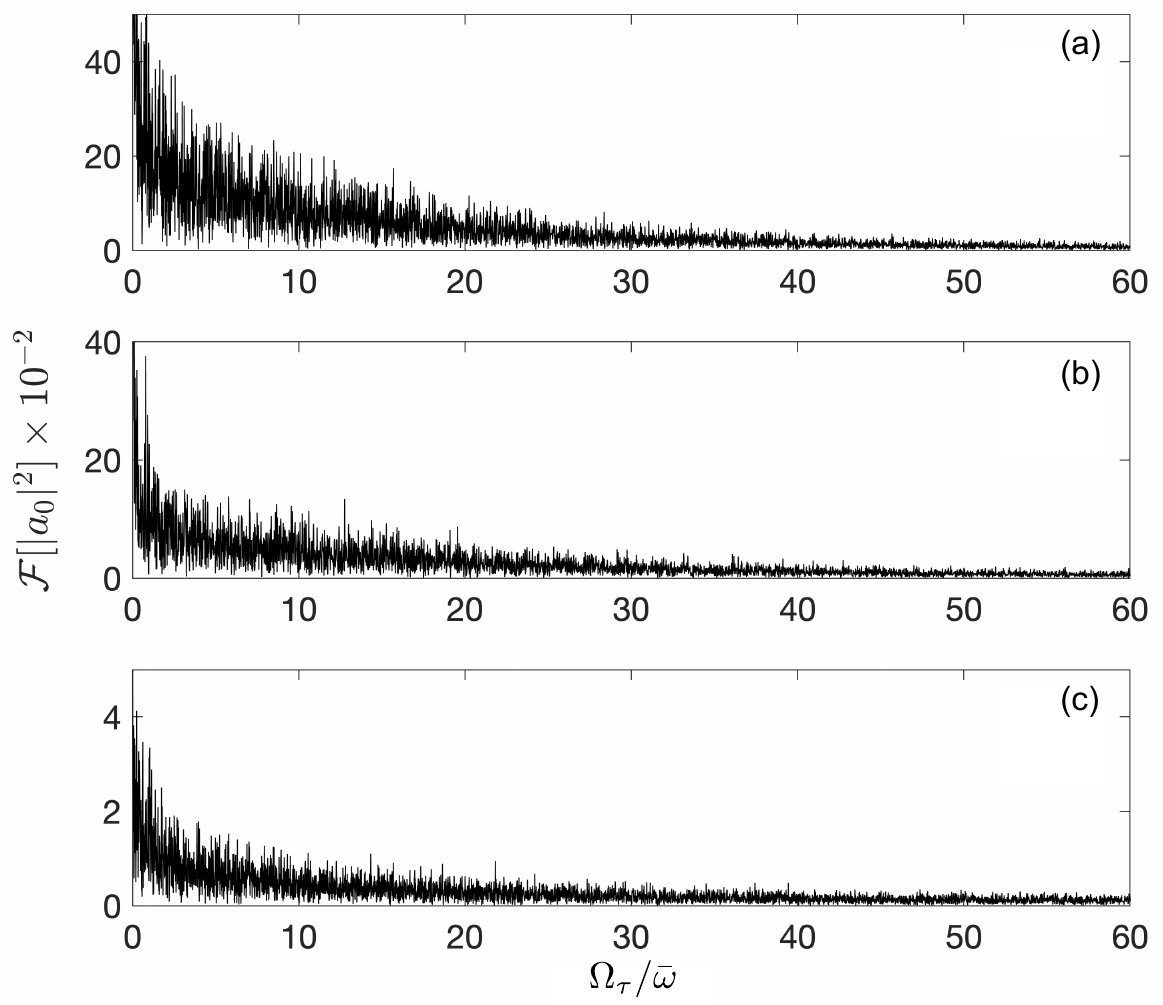}
\caption{{Power spectra of the population of the lowest energy harmonic oscillator mode for a single realization of the simulations of Fig.~\ref{fig:mom-distr} (a)--(c), respectively. We have used \(\Omega_{\tau}\) to denote the dimensionless Fourier frequency component of the dimensionless time $\tau$, and expressed it in units of the dimensionless trap frequency $\bar{\omega}$. It can be seen that all spectra are continuous. }}
\label{fig:powerspectrum}
\end{figure} 

This conclusion is in contradiction with our results for the quantum Newton's cradle.  Instead, we find that for the parameters of our simulations the system is not quasi-integrable, and that it does relax to a higher energy state that appears to be in thermal equilibrium.  To further support our findings, we have performed the same power spectrum analysis as Bland~\emph{et al.} for the Newton's cradle collisional dynamics of Fig.~\ref{fig:mom-distr}. The power spectrum is determined by calculating the amplitude of the harmonic oscillator ground state $\phi_0(x)$ making up the classical field $\psi(x,t)$ of a single trajectory
\begin{equation}
    a_0(t) = \int dx \phi_0^*(x) \psi(x,t),
\end{equation}
and  then calculating the Fourier transform of the population $|a_0(t)|^2$.  The results are shown in Fig.~\ref{fig:powerspectrum}, where it can be seen that the power spectra of the simulations are essentially continuous, rather than quasi-discrete as was the case for Bland~\emph{et al.}~\cite{Bland_2018}.  We have established that this difference is due the fact that Bland~\emph{et al.}\ only considered systems with relatively weak nonlinearities, such that the power spectrum is not strongly broadened from the integrable noninteracting (ideal Bose gas) case.  We have reproduced the results of Bland~\emph{et al.}, and we find that as the normalization of the initial state is increased, the quasi-discreteness of the power spectra for the 1D harmonic trap in their setup turns into a continuous spectrum.  We present these results in full in Appendix~\ref{A}.  

In support of our findings, we also note the recent work by Bastianello~\emph{et al.}~\cite{Bastianello_2020}, who consider the relaxation of a 1D harmonically trapped Bose gas using a generalized hydrodynamics approach. They simulated a related quantum Newton's cradle experiment in a 1D quasicondensate that was initialised in a longitudinal double-well potential and then released into a single harmonic well \cite{Bouchoule_QNC}. Bastianello~\emph{et al.}\ also find that the system thermalizes at long times. This is in agreement with our conclusions, rather than those of  Bland \emph{et al.}~\cite{Bland_2018}.
}

\subsection{Candidate dynamical regimes}
\label{phase-diagram}

The qualitatively different types of dynamical behaviour illustrated in Sec.~\ref{examples} ultimately depend on the pair of intrinsic dimensionless parameters of the initial equilibrium quasicondensate, ($\chi_0, \bar{\omega}$), and the Bragg momentum $2\hbar k_0$ imparted onto each half of the split cloud. The interplay between these three parameters can, in the first instance, be analyzed using simple geometric considerations. This will allow us to construct a qualitative overview of the expected different types of dynamical behaviour, which can be broadly classified as (I) aperiodic---displaying fast thermalization; (II) quasiperiodic---displaying intermediate to slow thermalization timescales, and (III) periodic---in which case the Newton's cradle collisional oscillations persist for many oscillation periods and the thermalization is the slowest. 

The first of the simple geometric considerations is the requirement that in order to set the initial quasicondensate into a well-defined Newton's cradle collisional regime and observe persisting oscillations over many periods, the two momentum components of the cloud must be well separated in momentum space. This means that the difference between the Bragg momenta $\pm 2\hbar k_0$ must be much larger than the characteristic momentum width of the cloud, which we denote via $\sigma_k$:
\begin{align}
4 k_{0} \gg \sigma_{k}.
\label{eqn:unsep_mom}
\end{align} 

The second consideration is that, even if the initial classical field components are well separated in momentum space, the two clouds will relax quickly, within the a few oscillation cycles if the respective position-space density distributions remain largely overlapping
and thus do not separate well in position-space. Thus, the second requirement for setting the initial quasicondensate into a a well-defined Newton's cradle collisional regime is that the maximum separation of the clouds $x_{\max}$ is much larger than the characteristic width of the cloud in position space, $\sigma_x$: $x_{\max}\gg \sigma_x$. Given that for a simple harmonic motion with a maximum momentum $2\hbar k_0$, the maximum displacement is given by $x_{\max}=2\hbar k_0/m \omega$, this condition can be approximated via
\begin{align}
2k_{0} \gg \sigma_{x}/l_{ho}^2, 
\label{eqn:f-q}
\end{align}

Equations (\ref{eqn:unsep_mom}) and (\ref{eqn:f-q}) can be further simplified and rewritten in terms of our dimensionless parameters if we approximate the characteristic size of the quasicondensate $\sigma_x$ by the TF radius, $R_{TF}=\sqrt{2\mu/m\omega^2}$ and the momentum width of the cloud by the inverse of the temperature-dependent phase coherence length in the trap centre \cite{bsdk16}, $\sigma_k\simeq1/l_{\phi}$, with $l_{\phi}=\hbar^2 \rho_0/mk_BT$. Introducing the dimensionless Bragg momentum $q_0\equiv k_0x_0$, Eqs.~(\ref{eqn:unsep_mom}) and (\ref{eqn:f-q}) can now be rewritten, respectively, as:
\begin{align}
q_0\gg & \frac{1}{4\bar{\mu}}=\frac{\chi_0^{2/3}}{4}, 
\label{eqn:unsep_mom_dimensionless} \\
q_0\gg & \sqrt{\frac{\bar{\mu}}{2}}=\frac{1}{\sqrt{2}\,\chi_0^{1/3}}, 
\label{eqn:f-q_dimensionless}
\end{align}
where we note that the quasicondensate regime requires $\chi_0\ll1$, and the condition that the density profile is well approximated by the TF inverted parabola, $R_{TF}\!\gg \!l_{ho}$, is equivalent to $\bar{\omega}\!\ll \!\bar{\mu}\!=\!\chi_0^{-2/3}$.

\begin{figure}[tbp]
\includegraphics[width=8.5cm]{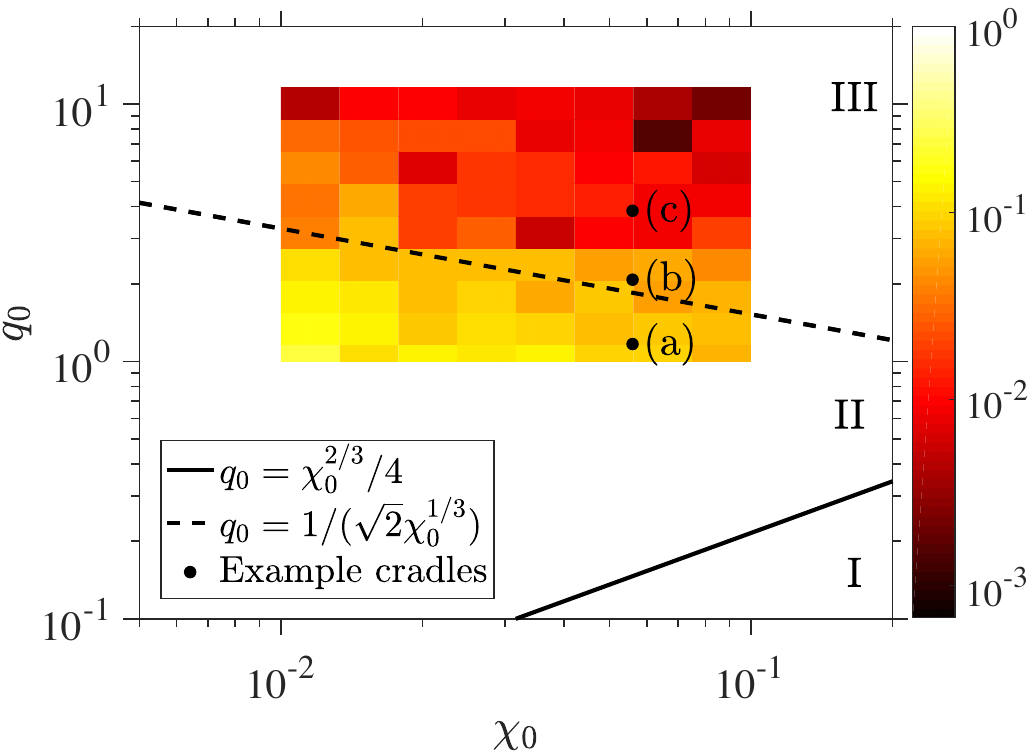}
\caption{Diagram of the dynamical regimes for the quasicondensate Newton's cradle in the $\chi_0$ versus $q_0$ parameter space, with $\bar{\omega}=0.0696$. The crossover boundaries (solid and dashed lines) between the different regimes, corresponding to fast (regions I and II) and slow (region III) relaxation, equate to the right-hand-sides of Eqs.~(\ref{eqn:unsep_mom_dimensionless}) and (\ref{eqn:f-q_dimensionless}). The circles labelled (a), (b) and (c) represent, respectively, the parameter combinations used in the examples displayed in Figs.~\ref{fig:mom-distr}, \ref{fig:fits}, and \ref{fig:decay} (see text). The coloured rectangle corresponds to the scanned parameter space of $\chi_0$ and $q_0$, for which we extracted the dimensionless decay rate $\overline{\Gamma}$ from the numerical simulations (see Sec.~\ref{decay}). As we see, the crossover boundary between the regions II and III predicted from simple arguments is in good agreement with the quantitative picture emerging from the simulations of the decay rate.
}
\label{fig:phase-diagram}
\end{figure}

Using the right-hand-sides of Eqs.~(\ref{eqn:unsep_mom_dimensionless}) and (\ref{eqn:f-q_dimensionless}) as crossover boundaries of different types of behaviour we can now construct a candidate diagram of the dynamical regimes (see Fig.~\ref{fig:phase-diagram}) for observing different relaxation scenarios of the quasicondensate Newton's cradle. This will be verified numerically in the next section. In the regions I and II we expect fast thermalization due to the fact that the two momentum components of the split quasicondensate do not separate well in momentum (region I) or in position spaces (region II); the example in Fig.~\ref{fig:mom-distr}(a) corresponds to conditions from region II. Region III corresponds to quasiperiodic behaviour and intermediate thermalization timescales, and this is when, despite the fact that the two components are well separated in both momentum and position spaces, the dephasing during the first few collisions due to the strong nonlinearity acts as a strong perturbation to persistent periodic behaviour; the examples (b) and (c) in Fig.~\ref{fig:mom-distr} corresponds to conditions from region III. 

\subsection{Characteristic relaxation rate}
\label{decay}

The observable we use to characterize the relaxation rate of the system is the rms momentum width, for the momentum distribution averaged over each ($m$th, $m=1,2,3,...$) period of Newton's cradle collisional oscillations:
\begin{align}
W_m &= \sqrt{\frac{\int dk\, k^2 \, n_m(k)}{\int dk \, n_m(k)}},
\end{align}
where 
\begin{align}
n_m(k) &= \frac{1}{T_\omega} \int_{T_\omega\times (m-1)}^{T_\omega\times m} dt \, n(k,t),
\end{align}
is the momentum distribution averaged over the \(m\)th oscillation and $T_{\omega}=2\pi/\omega$ is the oscillation period. Here, averaging over an oscillation period separates the oscillatory dynamics and the internal structure of instantaneous momentum distributions from the gradual net relaxation and the approach of \(W_m\) to a stationary value.

\begin{figure}[tbp]
\includegraphics[width=8.5cm]{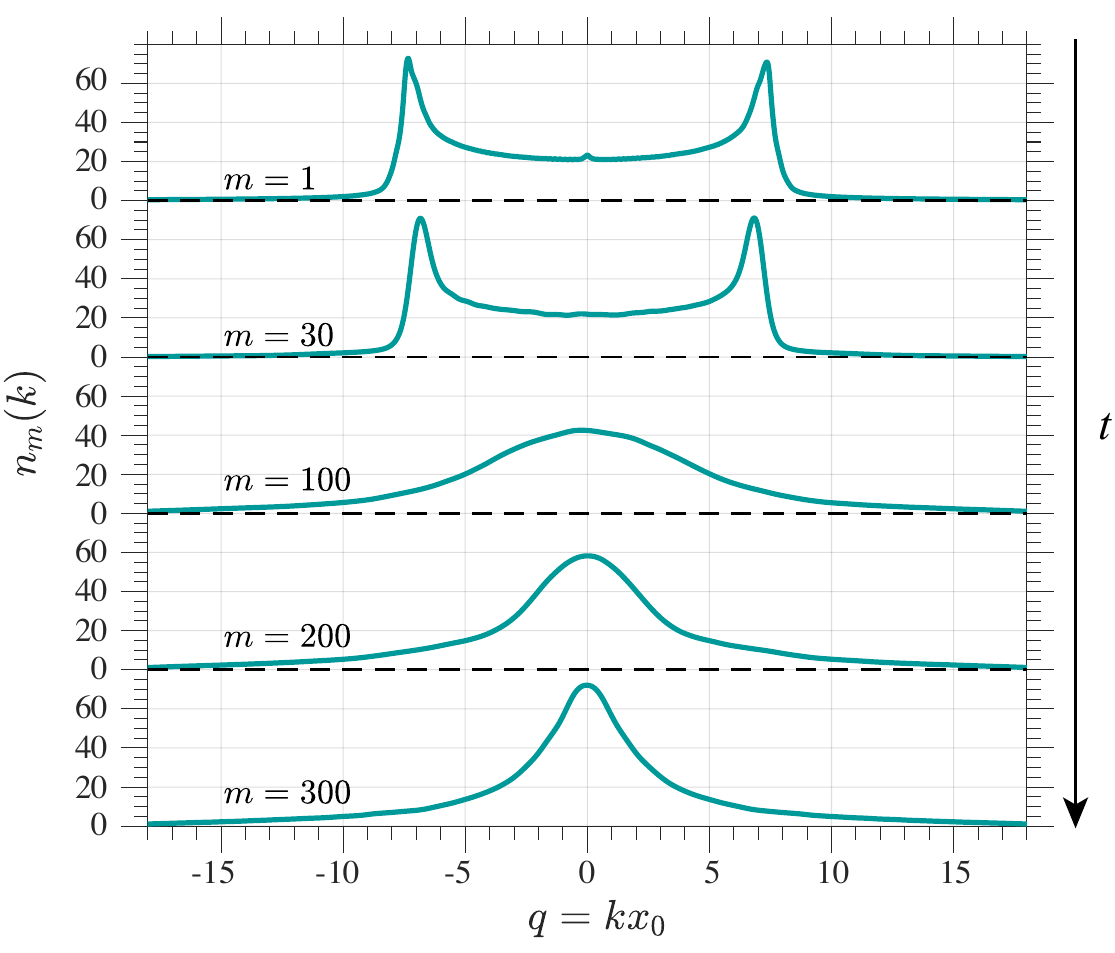}
\caption{Momentum distributions averaged over one ($m$th) oscillation period, $n_m(k)$, for parameter values corresponding to \((\chi_0,q_0,\w)=(0.0562,3.67,0.0696)\), \emph{i.e.}, example (c) in Fig.~\ref{fig:mom-distr}. Distributions for $m\!=\!1,30, 100, 200,$ and $300$ are shown. The average distributions for the earlier oscillations, \(m=1\) and \(30\), still have two separate peaks at the initial momentum kicks \(\pm 2q_0\) and a smaller third peak at $q=0$, indicating there is still some oscillatory dynamics present in the system, while the large-$m$ averages approach stationary thermal distributions as those in Fig.~\ref{fig:fits}.
}
\label{fig:nm}
\end{figure}

\begin{figure}[tbp]
\includegraphics[width=8.3cm]{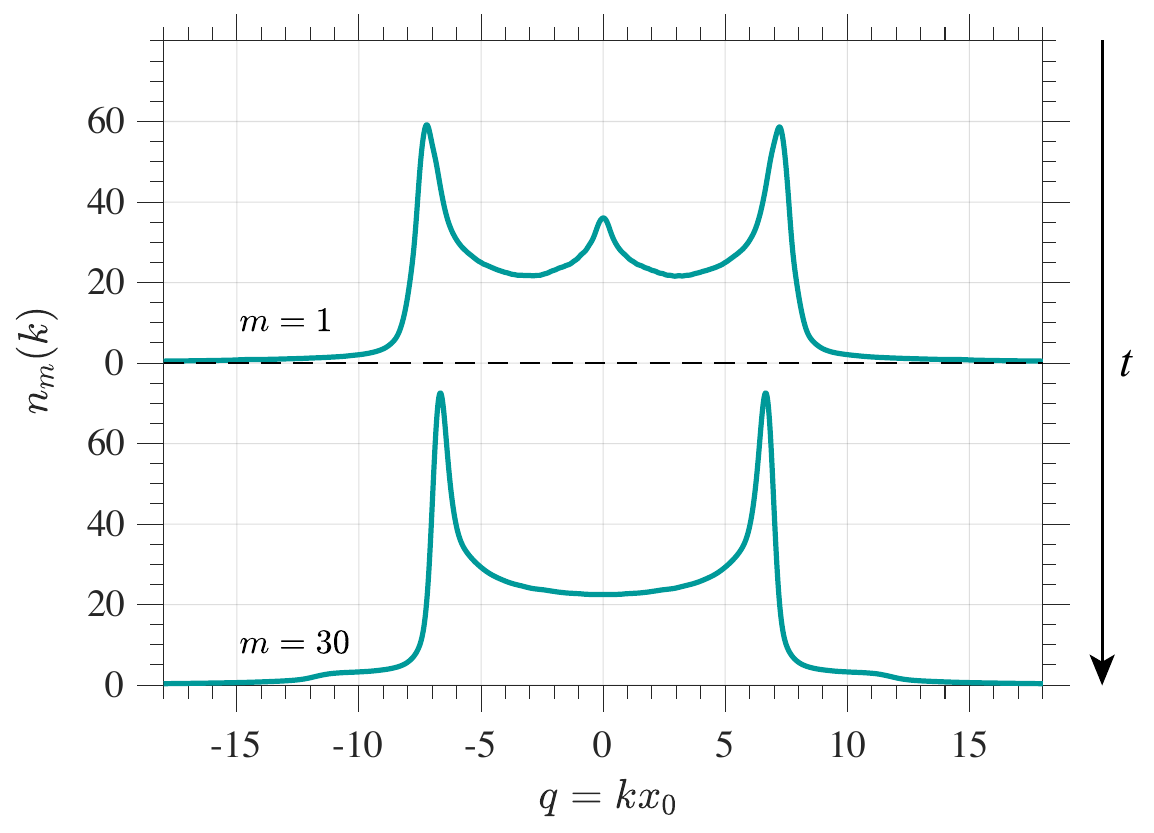}
\caption{Momentum distributions averaged over one ($m$th) oscillation period, $n_m(k)$, for parameter values corresponding to \((\chi_0,q_0,\w)=(0.0562,3.67,0.0696)\) as in Fig.~\ref{fig:nm}. However, here we initialize the system with ten percent of particles remaining unperturbed by the Bragg pulse (see text). We see that in this case the distinct three peaked distribution, observed in Refs.~\cite{kww06,agbk16}, does occur initially (\(m=1\)), in contrast to Fig.~\ref{fig:nm}, however, the three peaked structure eventually dampens out as can be seen from the example for \(m=30\).
}
\label{fig:zero_peak}
\end{figure}

Examples of the momentum distributions $n_m(k)$ averaged over the $m$th oscillation period, for $m=1,30, 100, 200, 300$ and the parameter values of Fig.~\ref{fig:mom-distr}(c), are shown in Fig.~\ref{fig:nm}.
It is interesting to see that the three peaked structure present in the equivalent averages found experimentally in Ref.~\cite{kww06} and theoretically in Ref.~\cite{agbk16} are not present in Fig.~\ref{fig:nm}. In order to explain this discrepancy we performed a simulation with equivalent parameters to those in Fig.~\ref{fig:nm}, however rather than applying an idealized Bragg pulse of Eq.~(\ref{Bragg}) we initialize the system with ten percent of particles remaining in the zero momentum state, \tit{i.e.}, unperturbed. This more closely resembles the physical Bragg pulse implemented in Refs.~\cite{kww06,agbk16}. As can be seen in Fig.~\ref{fig:zero_peak}, using this initialization the three peaked distribution does occur initially, implying that this feature occurs experimentally due to the imperfect nature of the Bragg pulse.

To extract the relaxation rate \(\Gamma\) from the $c$-field simulation results, we fit an exponential decay of the form
\begin{align}
W_m &= A \exp \left(-\overline{\Gamma} m \right)+B,
\label{eqn:decay}
\end{align}
to the numerically computed values of \(W_m\). Here, $\overline{\Gamma}=\Gamma T_\omega$ is the dimensionless relaxation rate, \(m\) is the oscillation number, and the numerical constants $A$ and $B$ fix the initial and final (relaxed) values of $W_m$. 

\begin{figure}[tbp]
\includegraphics[width=8.5cm]{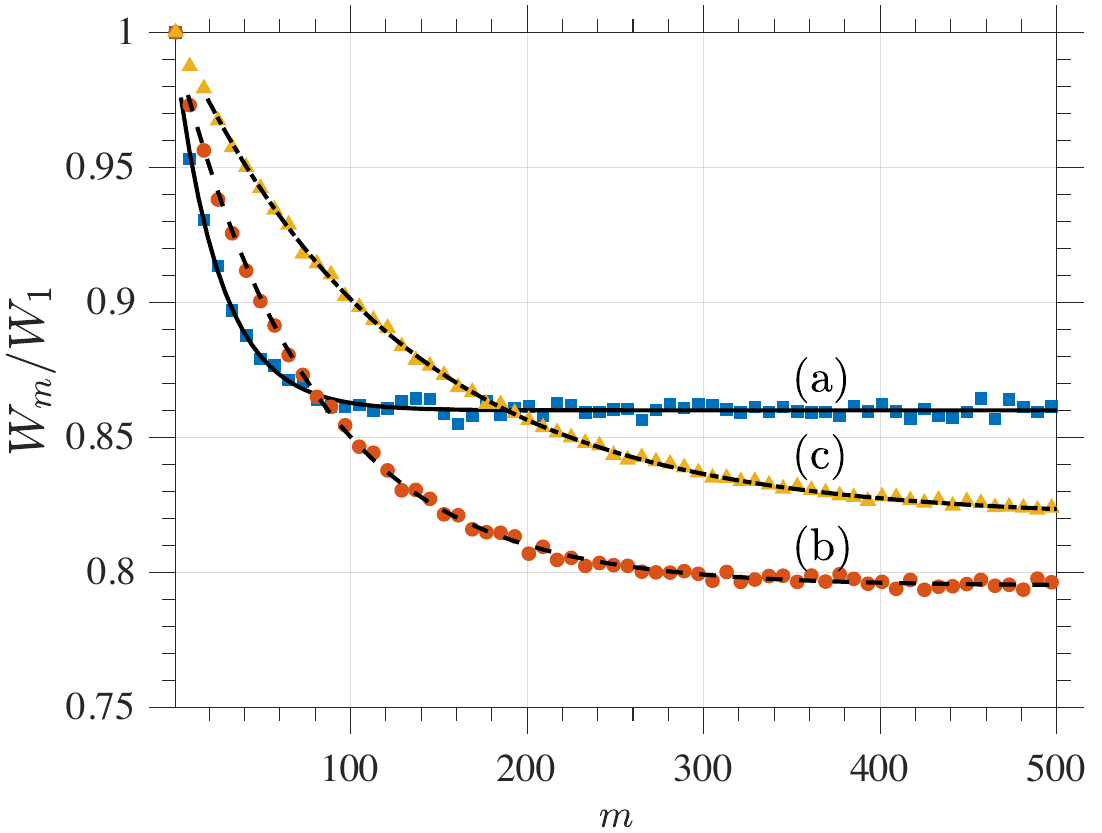}
\caption{Relaxation of the rms width \(W_m\) of the average momentum distribution over the $m$th oscillation period as a function of oscillation number \(m\). The symbols correspond to the numerically evaluated values of the width $W_m$, for the same parameters as in Fig.~\ref{fig:mom-distr}, whereas the solid, dashed, and dash-dotted lines are based on exponential fits of Eq.~(\ref{eqn:decay}); the widths are rescaled by their respective values of \(W_1\) in order to be easily visible on the same graph. Note that the exponential fits begin after the respective prethermalization periods which are are not well approximated by a single exponential. The long-time relaxation rates $\overline{\Gamma}$ extracted from these fits are as follows: (a) $\overline{\Gamma}=0.06$; (b) $\overline{\Gamma}=0.025$; and (c) $\overline{\Gamma}=0.0065$. 
}
\label{fig:decay}
\end{figure}

In Fig.~\ref{fig:decay}, we plot examples of relaxation of $W_m$ for parameter values corresponding to Fig.~\ref{fig:mom-distr}(a--c). 
As we can see, after the initial fast pre-thermalization stage \cite{bbw04,cr10,kwe11,kitagawa2011dynamics,gklkrsmsds12,smith2013prethermalization} during which the phase dynamics damps out, the subsequent decay of the rms width $W_m$ towards the final thermal distribution, characterized in Fig.~\ref{fig:fits}, is well approximated by an exponential.

By repeating the fitting procedure for a range of parameter values of $(\chi_0, q_0,\bar{\omega})$ and extracting the respective relaxation rates $\overline{\Gamma}$, we can verify the proposed nonequilibrium dynamical regimes of Fig.~\ref{fig:phase-diagram}. This is illustrated through the colour plot embedded in Fig.~\ref{fig:phase-diagram}, where we show a density plot of $\overline{\Gamma}$ over $(\chi_0, q_0)$ parameter space for a fixed value of $\bar{\omega}=0.0696$. As we see from these results, for an experimentally feasible range of parameters, the relaxation rate $\overline{\Gamma}$ typically ranges within $\overline{\Gamma} \in [0.002, 0.2]$, and the crossover boundary between the regions II and III, evident at $\bar{\Gamma}\sim0.1$, is in good agreement with the boundary predicted in Sec.~\ref{phase-diagram}.

Taking the trap frequency of $\omega/2\pi= 150$ Hz as an example (corresponding to the values used in the original quantum Newton's cradle experiment in a strongly interacting 1D Bose gas \cite{kww06}), this range of the dimensionless $\overline{\Gamma}$ corresponds to $\Gamma \in [0.3, 30]$ s$^{-1}$, and hence the characteristic relaxation time for the quasicondensate Newton's cradle is expected to be in the range of 0.033 to 3.3 seconds, depending on the actual interaction strength and temperature of the gas. For comparison, for $\omega/2\pi= 5$~Hz, which is the typical value used in the coherently split quasicondensate experiments \cite{hlfss07}, our results convert approximately to typical relaxation times of $1$ to $100$ seconds. This is at least three orders of magnitude larger than the millisecond time scale of relaxation of phase dynamics in a 1D quasicondensate \cite{hlfss07}. At the same time, this is comparable to the relaxation times scales observed in the Newton's cradle setup of a strongly interacting 1D Bose gas. We thus predict that for an equivalent quantum Newton's cradle experiment in a weakly-interacting 1D Bose gas the relaxation rate is of similar order of magnitude as in the strongly interacting case.

\section{Conclusion}
\label{conclusion}

In conclusion, we studied the collisional dynamics of a harmonically trapped weakly-interacting 1D quasicondensate in the same quantum Newton's cradle setting as the experiment performed by Kinoshita \tit{et al.} \cite{kww06} in the strongly interacting (near-Tonks-Girardeau) regime. By parametrizing the system in terms of just two (for any given trap frequency) dimensionless parameters, $\chi_0$ and $q_0$, which encompass the interaction strength, the temperature of the gas, and the magnitude of the Bragg momentum, we identify the different dynamical regimes of the system in terms of its relaxation rate to the final equilibrium state. 

We find that the final relaxed state of the system can be well characterized by a thermal distribution, where the kinetic energy imparted to the system by the initial Bragg pulse is distributed amongst the internal degrees of the system and results in a higher equilibrium temperature of the final state. Relaxation to this thermal state, rather than to a state which would require characterization via a generalized Gibbs ensemble, implies that harmonic confinement here cannot be regarded as a weak perturbation from integrability of the respective uniform 1D Bose gas. Accordingly, a 1D quasicondensate in this nonequilibrium scenario does not offer itself as a system to which a quantum analog of the KAM theorem could be applied.

The characteristic relaxation timescales are predicted to be of the order of tens of seconds for typical experimental parameters, which in turn correspond to hundreds of repeated collisions taking place before equilibrium is reached. This is similar to, and only an order of magnitude shorter than, the typical relaxation timescales observed in the strongly interacting quantum Newton's cradle. The difference can be explained by the effect of repulsive interactions in the system, which strongly reduce the two-body correlations and hence suppress the two-body collisional rates in the Tonks-Girardeau regime.

Our system can be realized using currently available experimental techniques of creating equilibrium 1D Bose gases, except that one has to maintain an additional, dynamical 1D condition of $\hbar^2(2k_0)^2/2m \ll \hbar\omega_{\perp}$, which is required for eliminating transverse excitations of the gas due to the (large) longitudinal kinetic energy imparted onto the atoms by the Bragg pulse and hence could be challenging. The nonequilibrium scenario that we studied represents a directly comparable counterpart of the original quantum Newton's cradle setup, except that the collisional dynamics now takes place in a weakly interacting gas. A similarly comparable scenario, which could provide further insights into thermalization in strongly and weakly-interacting 1D Bose gases, would be to perform Hofferberth \tit{et al.}  experiment \cite{hlfss07} in the strongly interacting regime.

\begin{acknowledgments}
This research was supported by the Australian Research Council (ARC) through ARC Discovery Project grant DP170101423 (K.V.K.) and the ARC Centre of Excellence for Engineered Quantum Systems  (project ID CE170100009) (M.J.D.).  
 \end{acknowledgments}

\appendix 
\section{Spectral analysis of the 1D GPE in the harmonic oscillator basis}
\label{A}

{
In this Appendix we reproduce the results of Bland~\emph{et al.}~\cite{Bland_2018} for randomized initial conditions.  The Galerkin expansion used in the simulations of Bland~\emph{et al.}~\cite{Bland_2018} is identical to using a restricted harmonic oscillator basis for numerically solving the projected Gross-Pitaevskii equation in 1D.
We find that the conclusions drawn based on the results reported by Bland~\emph{et al.}\ cannot be generalized to systems with effective nonlinearities much larger than the ones considered in Ref.~\cite{Bland_2018}.  We demonstrate a transition from quasi-discrete to continuous power spectra with increasing normalisation of the field.

\begin{figure}
\includegraphics[width=8.5cm]{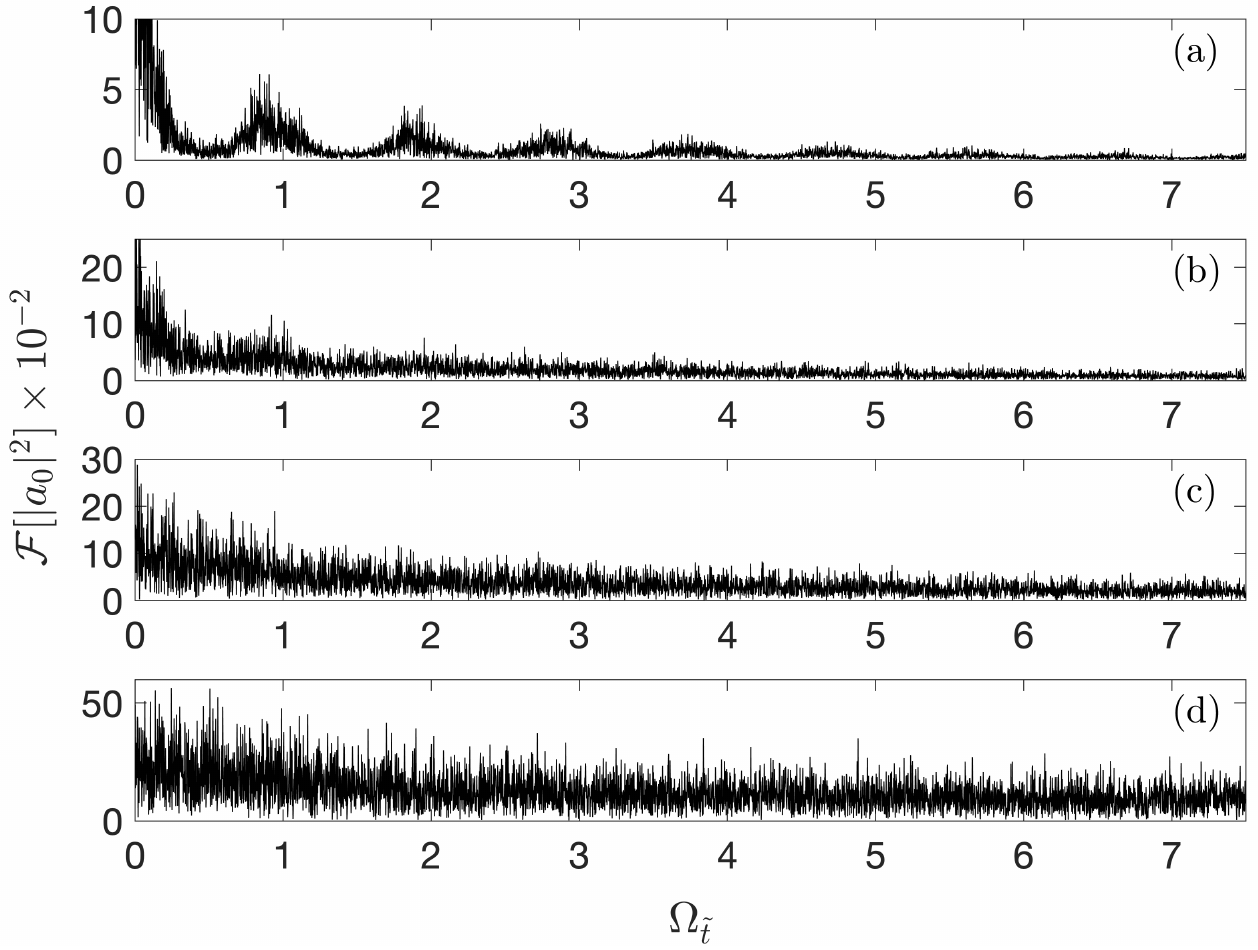}
\caption{{Comparison of the power spectra of $|a_0(t)|^2$ for GPE simulations with $M=16$ for increasing normalization, corresponding to the infinite temperature simulations  as described by Bland~\emph{et al.}~\cite{Bland_2018}. Note we have used \(\Omega_{\tilde{t}}\) to denote the corresponding Fourier dimension of the dimensionless time \(\tilde{t}\). The normalizations are (a) $N=15$, (b) $N=30$, (c) $N=50$, and (d) $N=150$. For $N=15$ the spectra is quasi-discrete, but  for larger values of $N$ the spectra are continuous.}}
\label{fig:many_different_N}
\end{figure}

The conclusion of Bland~\emph{et al.}~is that the quasi-integrability---as observed in simulations of dark soliton oscillations in a 1D GPE in a harmonic trap--- is due to the observation that (from Bland~\emph{et al.} abstract): \emph{``the finite-mode dynamics always produces a quasi-discrete power spectrum, with no visible continuous component, the presence of the latter being a necessary manifestation of ergodicity. This conclusion remains true when a strong random-field component is added to the initial conditions.''} This conclusion is based on performing two sets of simulations:
\begin{enumerate}
\item Oscillations of a dark soliton in a harmonic trap, using both the full GPE, as well as the $M=16$ mode Galerkin truncation.  The power spectra of the first and second Galerkin amplitudes are found to be discrete (Figure 2, left column, of Ref.~\cite{Bland_2018}).
\item The dynamics of an initial condition with random amplitude and phase, for both the $M=16$ mode Galerkin truncation as well as the full GPE.  Bland~\emph{et al.} claim that the results of their power spectra for the first Galerkin component in their Fig.~3 for the harmonic trap are sufficiently similar to the discrete case that the conclusion can be generalized, \textit{i.e.} quasi-integrable behaviour should always occur for the full 1D GPE in a harmonic trap.
\end{enumerate}

In this Appendix we focus on the results and conclusions drawn from the second set of simulations with randomized initial conditions, as these are most relevant for our manuscript on the thermalization of Newton's cradle for 1D quasicondensates. We  numerically solve the dimensionless GPE
\begin{align}
    i \frac{\partial \psi}{\partial \tilde{t}} &= -\frac{1}{2} \frac{\partial^2 \psi}{\partial \tilde{x}^2} + |\psi|^2 \psi + \frac{1}{2} \tilde{\omega}^2 x^2 \psi,
    \label{eqn:gpe_bland}
\end{align}
as in Ref.~\cite{Bland_2018}, where \(\tilde{x} = x/l_h\) \(\tilde{t} = t \mu / \hbar \), and \(\psi = \sqrt{2 |a_s|m \omega /\hbar^2} \Psi\) are the dimensionless quantities with \(l_h = \hbar/\sqrt{m \mu}\) being the healing length. As we fix \(\tilde{\omega} = 1\), the only free parameter in Eq.~(\ref{eqn:gpe_bland}) is the normalization of the dimensionless field \(N= \int |\psi|^2 d \tilde{x} \). In the Thomas-Fermi approximation we have 
\begin{align}
    N \approx \frac{8}{3} \frac{1}{\chi_0^{2/3} \bar{\omega}} = \frac{8 \mu}{3 \hbar \omega}.
\end{align}
In this appendix we integrate the GPE using a Galerkin method and represent $\psi$ on a finite basis of harmonic oscillator basis states. We have performed simulations with $M=16$ modes and the same randomized initial conditions, but with increasing normalisation.  Our results show that quasi-integrable behaviour does not occur for systems sizes much larger than those corresponding to the normalization of $N=15$ considered by Bland \emph{et al}.

\begin{figure}
\includegraphics[width=8.3cm]{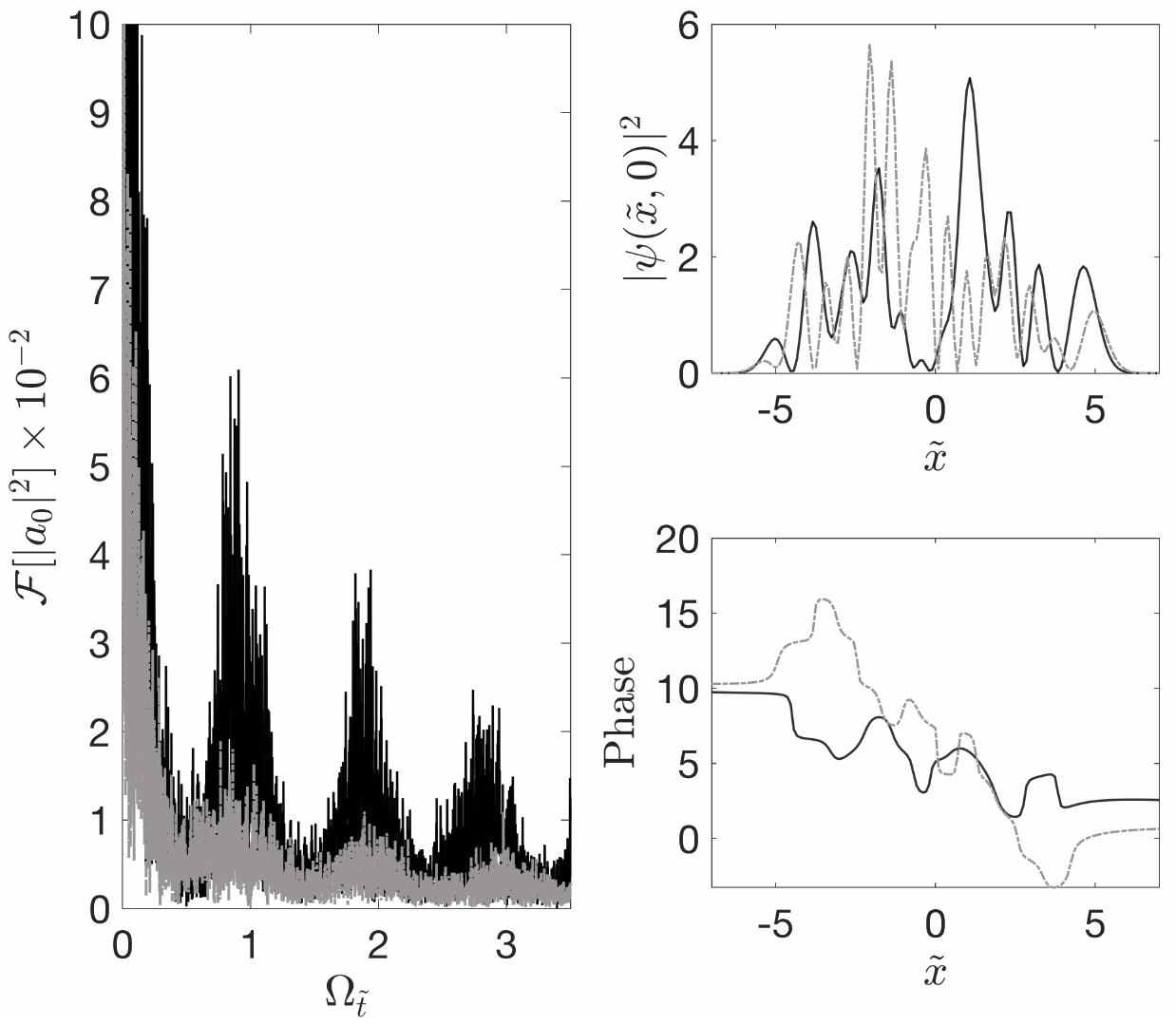}
\caption{{Comparison of different randomized initial conditions for the same normalization of $N=15$, with $M=16$.  Left: Comparison of the power spectra of $|a_0(t)|^2$  for two different randomized initial conditions (black and grey lines) as described by Bland~\emph{et al.}, with normalization $N=15$. Right: Comparison of the respective initial densities (top) and phases (bottom).}}
\label{fig:randomIC}
\end{figure}

In Fig.~\ref{fig:many_different_N} we reproduce the results presented in Fig.~3 of Bland \emph{et al.}~for randomized initial conditions with $M=16$ modes but with varying normalization. As in the main text,  the power spectrum is determined by calculating the amplitude of the harmonic oscillator ground state $\phi_0(x)$ making up the classical field $\psi(x,t)$ of a single trajectory
\begin{equation}
    a_0(t) = \int dx \phi_0^*(x) \psi(x,t),
\end{equation}
and  then calculating the Fourier transform of the population $|a_0(t)|^2$.  The results for the case of normalization $N=15$ corresponding to those of Bland \emph{et al.}, Fig.~\ref{fig:many_different_N}(a), do indeed show a quasi-discrete spectrum.  However,  doubling the normalization to $N=30$, Fig.~\ref{fig:many_different_N}(b), entirely changes the character of the spectra---there is no hint of discreteness. As the normalisation is further increased to $N=50$ and $N=150$, Fig.~\ref{fig:many_different_N}(c,d), it can clearly be seen that the quasi-discrete spectrum is no longer apparent. In comparison, the  equivalent normalization for the simulations presented in the main text of this paper is \(N \simeq 261\). 
We further note that since the random initial condition used in the Bland~\tit{et al.}\ simulation \cite{Bland_2018} is effectively at infinite temperature, the system being simulated is in the nearly classical ideal gas regime for sufficiently low \(N\) which can be analyzed using perturbation theory~\cite{kgds03,kgds05,Deuar_2009,jabkb11}. This implies that it is the proximity to the ideal Bose gas regime that is responsible for the apparent quasi-integrable behaviour in Band \tit{et al.} simulations. Increasing the interaction strength (or effectively the normalisation) leads, on the other hand, to the broadening of the power spectra and eventually to the continuous spectrum.

Finally, in Fig.~\ref{fig:randomIC} we demonstrate that even for $N=15$ the discreteness of the calculated power spectra can vary considerably for different samples of random initial conditions.  The spectrum of the grey curve in the left panel of Fig.~\ref{fig:randomIC} is almost continuous, whereas the black curve is quasi-discrete, yet both initial states were sampled in the same manner.  There is no easily discernible qualitative difference in the initial conditions of the field density and phase as shown on the right of the figure.

In summary, our numerical results show a continuous power spectra for the coefficient $|a_0(t)|^2$ for 1D GPE dynamics in a harmonic trap in 1D for sufficiently large nonlinearities.  This suggests that for the scenarios we consider in this paper we can expect ergodic dynamics and thermalization, in contrast to the conclusions drawn by Bland~\emph{et al.}~\cite{Bland_2018}.  These observations are also supported by by the work of Bastianello~\emph{et al.} \cite{Bastianello_2020}.}


\begin{thebibliography}{56}
\bibitem [{KKM()}]{KKMD}%
  \BibitemOpen
  \href@noop {} {}\bibinfo {howpublished} {This equivalence stems from a simple
  model in which the initial 1D quasicondensate at temperature $T$, in a
  harmonic trap of frequency $\omega$, is adiabatically split into two separate
  phase-coherent quasicondensates at half the density and the same
  characteristic phase coherence length as before. Approximating the density
  profile of each new quasicondensate by half of the initial Thomas-Fermi
  inverted parabola and the respective phase coherence length in the trap
  centre by $l^{(0)}_{\phi}= \hbar^2\rho(0)/mk_BT$ \cite{bsdk16}, where
  $\rho(0)$ is the peak density of the initial quasicondensate, one can show
  that the shape and the phase-coherence length of each of the new
  quasicondensates correspond to an equivalent equilibrium system at an
  effective temperature $T/2$, in a harmonic trap of frequency
  $\omega/\sqrt{2}$. Yet, each of these quasicondensates now resides in a trap
  of frequency $\omega$. This is equivalent to invoking a confinement quench
  protocol on a single equilibrium quasicondensate, in which the trap frequency
  is suddenly (at time $t=0$) changed from the value of $\omega/\sqrt{2}$ to a
  new value of $\omega$. As shown in Refs.~\cite{Fang14,bsdk16}, such a
  confinement quench protocol invokes breathing-mode oscillations occurring at
  the frequency of $\sqrt{3}\,\omega$.}\BibitemShut {Stop}%
\bibitem [{con()}]{condition-position-space}%
  \BibitemOpen
  \href@noop {} {}\bibinfo {howpublished} {This condition can be contrasted to
  $g\rho \!<\!k_BT\!<\! \sqrt{\gamma} \hbar^{2} \rho^{2}/m$
  \cite{kgds03,kgds05}, which defines the quasicondensate regime if one is
  concerned with the position-space correlations and density-density
  fluctuations. In this case, at temperatures $k_BT\!<\!g\rho$, the
  position-space correlations become dominated by vacuum rather than thermal
  fluctuations, hence rendering the $c$-field method inapplicable.}\BibitemShut
  {Stop}%
\bibitem [{IBG()}]{IBG}%
  \BibitemOpen
  \href@noop {} {}\bibinfo {howpublished} {In the opposite side of the
  quasicondensate domain, \emph{i.e.}, for temperatures $k_BT>
  \sqrt{\gamma}\hbar^2\rho^2/m$, the gas crosses over to a highly degenerate
  nearly ideal {B}ose gas regime, where the interparticle interactions are only
  a small perturbation to the ideal {B}ose gas behaviour. While this regime can
  also be described by the $c$-field approach, as long as the gas is highly
  degenerate ($k_BT< \hbar^2\rho^2/2m$), such a description is beyond the scope
  of this paper.}\BibitemShut {Stop}%
\end{thebibliography}

%

\end{document}